1   **Host species and environmental effects on bacterial communities associated**

2   **with *Drosophila* in the laboratory and in the natural environment**

4   Fabian Staubach[1*], John F. Baines[2], Sven Kuenzel[2], Elisabeth M. Bik[3], and Dmitri A. Petrov[1]

5   **1** Department of Biology, Stanford University, 371 Serra St, Stanford, California, 94305-5020,

6   United States of America, **2** Max Planck Institute for Evolutionary Biology, August-

7   Thienemann-Str. 2, 24306  Plön, Germany, **3** Department of Microbiology & Immunology,

8   Stanford School of Medicine, 269 Campus Drive, Stanford, California, 94305, United States of

9   America

13  *corresponding author: Fabian Staubach,  Department of Biology, Stanford University, 371 Serra

14  St, Stanford, California, 94305-5020, United States of America, phone: +1 (650) 736 2249, fax:

15  (650) 723 6132, email: fabians@stanford.edu

17  **Abstract**

18  The fruit fly *Drosophila* is a classic model organism to study adaptation as well as the

19  relationship between genetic variation and phenotypes. Although associated bacterial

20  communities might be important for many aspects of *Drosophila* biology, knowledge about their

21  diversity, composition, and factors shaping them is limited. We used 454-based sequencing of a

22  variable region of the bacterial 16S ribosomal RNA gene to characterize the bacterial

23  communities associated with wild and laboratory *Drosophila* isolates. In order to specifically

24  investigate effects of food source and host species on bacterial communities, we analyzed

25  samples from wild *Drosophila melanogaster* and *D. simulans* collected from a variety of natural substrates, as well as from adults and larvae of nine laboratory-reared *Drosophila* species. We find no evidence for host species effects in lab-reared flies, instead lab of origin and stochastic effects, which could influence studies of *Drosophila* phenotypes, are pronounced. In contrast, the natural *Drosophila*–associated microbiota appears to be predominantly shaped by food substrate with an additional but smaller effect of host species identity. We identify a core member of this natural microbiota that belongs to the genus *Gluconobacter* and is common to all wild-caught flies in this study, but absent from the laboratory. This makes it a strong candidate for being part of what could be a natural *D.melanogaster* and *D. simulans* core microbiome. Furthermore we were able to identify candidate pathogens in natural fly isolates.

**Introduction**

Bacterial symbionts play important roles for metazoans covering the whole spectrum from beneficial mutualists to infectious, disease-causing pathogens. Benefits that hosts derive from mutualists are diverse and include extracting essential nutrients from food in humans [1], breaking down cellulose in Ruminantia [2], and light production by *Vibrio fisheri* in the light organs of the bobtail squid [3]. In arthropods, indigenous bacteria protect aphids from parasitoid wasps [4], protect beewolf larvae from infectious disease [5], and keep leaf tissue of fallen leaves photosynthetically active, providing larvae of leaf miner moths with nutrients [6]. Detrimental effects microbes have on their hosts range from lethal disease [7] to changing the sex ratio of the offspring in their favor [8].

Pathogens as well as mutualists not only interact with their hosts, but at the same time with other members of the often diverse host associated microbial community [9]. Indirect evidence for competition for ecological niches in the host comes from Staubach *et al.* [10] who found that the lack of the glycosyltransferase B4galnt2 in mice leads to the replacement of bacterial taxa by closely related taxa. Bakula [11] showed that *Escherichia coli* persists in *Drosophila* only when monoxenic and is quickly replaced by other bacteria upon exposure suggesting that there is competition between bacteria to colonize the fly. Ryu *et al.* [12] demonstrated that suppressing

54  the caudal gene by RNAi in *Drosophila* leads to replacement of an *Acetobacter* species by a

55  *Gluconobacter* species followed by strong pathological consequences. These examples indicate

56  that there is interaction and competition for ecological niches along the continuum of hosts and

57  microbes. Thus, a thorough understanding of host-microbe interactions also requires

58  comprehensive knowledge of host associated bacterial communities and the factors shaping

59  them.

60

61  These factors can roughly be grouped into two categories. The first category includes biotic and

62  abiotic environmental factors the host and its associated microbes are exposed to (e.g. diet). The

63  second category includes factors that are determined by host genetics. The relative importance of

64  these factors in shaping human associated microbial communities is a matter of recent debate

65  [13,14]. One approach to disentangle these effects is by studying the relationship of host genetic

66  divergence, diet, and divergence of microbial communities. A correlation of genetic divergence

67  between a set of host taxa and the divergence of their associated microbial communities would

68  suggest that genetic effects play a role in shaping these communities. On the other hand, a

69  correlation of microbial community composition with diet would suggest an effect of

70  environmental factors. This approach has been applied to a variety of mammals [15–17], but it

71  has proven difficult in mammals to control for diet and other environmental factors across host

72  taxa. Hence it is not yet clear, which factors are the strongest determinants of microbiota

73  composition.

74

75  In contrast to the complex microbial communities associated with mammals like humans and

76  mice, which are estimated to consist of hundreds or even thousands of taxa [10,18], some studies

77  suggest that only a handful of bacterial species dominate the microbial communities of

78  invertebrates [19,20]. This has turned a spotlight on *Drosophila* to serve as a simpler model for

79  understanding the complex interactions of hosts and their associated microbes [20–22]. The

80  *Drosophila* immune system is reasonably well understood [23] and the tractability of *Drosophila*

81  has helped to identify genes involved in specific interactions between host and microbes. This

82  includes genes underlying avoidance behavior towards harmful bacteria [24] and immune

3                                          

defense [25] as well as interactions with commensals [26] and beneficial bacteria that prevent pathogens from colonizing the host [12] or promote its growth [27,28].

As a first step in understanding the diversity of bacterial communities associated with *Drosophila* it is important to investigate flies under natural conditions. Most studies conducted to date focused on more specific interactions or those found in the lab [20,29], while few studies described the natural diversity of fly associated bacterial communities. Cox and Gilmore [30] included natural fly isolates and combined culture and culture-independent methods to characterize fly associated microbial communities. Corby-Harris *et al.* [31] focused their study on the diversity of microbial communities along latitudinal clines. Chandler et al. [32] conducted the most comprehensive analysis of bacteria associated with *Drosophila* by sampling a range of drosophilid flies from their natural food substrates. However, these studies were limited by either throughput or dependence on cultivation [33]. Although Chandler et al. [32] sampled flies from different natural substrates, their sampling scheme did not allow to directly disentangle host species and diet effects on the natural microbiota because this requires replicated, pairwise sampling of at least two host species from the identical substrate.

In order to understand bacterial communities associated with *Drosophila* and the factors shaping their diversity, we investigated the relative effects of food substrate and fly species. Accordingly, we analyzed *D. melanogaster* and *D. simulans* collected in pairs from different natural food sources, as well as under controlled lab conditions. Furthermore, we assessed the communities of nine lab-reared *Drosophila* species and their larvae to evaluate the influence of host genetic background on a broad scale. These species were selected to span the *Drosophila* genus and match the 12 species sequenced by Clark *et al.* [34] (*D. melanogaster*, *D. simulans*, *D. sechellia*, *D. yakuba, D. erecta, D. pseudoobscura, D. persimilis, D. virilis, D. mojavensis)*.

**Results**

In order to profile *Drosophila*-associated bacterial communities we amplified and sequenced ~300 bp (base pairs) of the 16S rRNA gene (see Materials and Methods) spanning the variable

112  regions V1 and V2. Three types of fly isolates were used in our study. The samples are listed in

113  Table 1. First, species-pairs of wild-caught *D. melanogaster* and *D. simulans* samples were

114  collected from different substrates (oranges, strawberries, apples, peaches, compost) at multiple

115  locations on the East and West Coast of the USA. Within each sample pair, *D. melanogaster* and

116  *D. simulans* individuals were collected at the same location, time, and substrate (mostly by

117  aspiration of individual flies from the same fruit), thereby controlling for environmental

118  variables to the extent possible in the field. This allowed us to study the effects of both, substrate

119  and host species on the composition of bacterial communities independently of each other.

120  Second, we included isofemale, wild-derived strains of *D. melanogaster* and *D. simulans* that

121  were reared in the Petrov lab for ~3 years after collection. Third, a variety of *Drosophila* species

122  from the UCSD Stock Center was chosen to complement the analysis. We primarily focused on

123  adults, but also studied bacterial communities in larvae of the lab-reared strains. We analyzed a

124  total of ~340,000 sequences that matched our quality criteria (see Materials and Methods).

125  ~130,000 sequences matched the *Wolbachia* 16S rRNA gene and were excluded from the

126  analysis. For Petrov lab *D. simulans* sample 6 , removal of *Wolbachia* sequences led to a very

127  low number of remaining sequences (18 sequences). Therefore, we excluded this sample from

128  further analysis (Supplementary Table 1 lists the total number of sequences and the proportion of

129  *Wolbachia* sequences for each sample).

130

131  <u>Diversity of bacterial communities associated with *Drosophila*</u>

132  For assessing the *Drosophila* associated bacterial diversity in general, we grouped all sequences

133  into 97% identity operational taxonomic units (OTUs) and calculated inverted Simpson diversity

134  indices [35]. Rarefaction curves are plotted in Figure 1. Bacterial communities associated with

135  lab-reared flies are strikingly less diverse than those of wild-caught flies ($P = 2.9 \times 10^{-5}$,

136  Wilcoxon test on Simpson diversity index), indicating a bias towards a few dominant species in

137  the lab compared to more complex and species-rich communities of wild-caught flies. However,

138  substantial variance of community diversity was found between individual samples from lab-

139  reared flies. While bacterial diversity in 14 out of 20 lab-reared fly samples is lower than in all

140  wild-caught samples, the diversity of lab-reared *D. erecta*, *D. persimilis*, *D. sechellia*, *D. virilis*,

and Petrov lab *D. melanogaster* sample 3 (m.pet3 in Table 1) lies within the range of wild-caught samples. The diversity observed in Petrov lab *D. melanogaster* sample 6 (m.pet6) is even higher than in wild-caught flies and its community composition appears to differ from the other Petrov lab samples (Figure 2C). Because this sample was unusual, we conducted all of the subsequent analyses with and without this sample, but did not notice any qualitative differences (data not shown). All of the analyses described below that include lab-reared samples also include this sample.

Comparing estimates of species richness and diversity from our study to estimates from lab-reared flies in Wong *et al.* [20] supports the notion that bacterial communities of lab-reared flies are less species rich (Table 2). Our species richness estimates from wild-caught flies are more than twice as high on average (43 vs 19, *P* < 0.001), if we exclude all OTUs that contain fewer than 10 sequences from our data as in Wong *et al.* [20]. Bacterial diversity, as measured by Shannon's diversity index, is also significantly higher in wild-caught flies (*P* < 0.01) from this study. We also compared the bacterial community diversity in this study to that observed in previous studies of wild-caught *Drosophila* bacterial communities, namely Corby-Harris *et al.* [31], Cox and Gilmore [30], and Chandler *et al.* [32]. A comparison of diversity indices among studies is provided in Table 2. Estimated species-richness is more than seven times higher in our study compared to all other studies (*P* < 0.001, Student's T-test). However, limiting our data artificially to 100 sequences per sample, which is well within the range of the sequencing depth of the above studies, results in an average Chao's richness estimate of 22 species. This is not significantly different from the richness estimates of the other studies on wild-caught flies, implying that different sequencing depths are responsible for the different species richness estimates. When we limit our sample size to 100 sequences to make our study more comparable to the clone library data from Cox and Gilmore [30] and Chandler *et al.* [32] we find values for Shannon's diversity index that are similar and even a bit higher (*P* < 0.001, Student's T-test) in these two studies. Note that direct comparison of diversity between studies is difficult due to different sample preparations (whole flies, fly guts, washing procedure), sequencing depths, and different regions of the bacterial 16S rRNA gene that were used for the analysis (see Table 2).

170

Bacterial community composition

In order to examine which bacterial taxa are associated with *Drosophila*, we classified the 16S

rRNA gene sequences by aligning them to the SILVA reference database [36] using MOTHUR

[37]. The results are summarized in Figure 2. Our results show that, on the family level, the

combined communities are dominated by Acetobacteraceae (55.3%) and Lactobacillaceae

(31.7%) (Figure 2A). Leuconostocaceae (3.8%), Enterobacteriaceae (3.3%) and Enterococcaceae

(1.9%) are less abundant. All five of these families are known to be associated with *Drosophila*

[6,32] including certain *Drosophila* pathogenic *Enterococcus* strains. The remaining sequences

(~3.9%) are low abundance families mainly belonging to the Proteobacteria.

180

In addition to the differences in overall diversity described above, different bacterial genera dominate the communities of lab-reared and wild-caught flies (Figure 2B). The dominant genera also vary sharply between flies from the Petrov lab and the UCSD Stock Center. Specifically, communities associated with wild-caught flies are dominated by *Gluconobacter* (39.3% average relative abundance), *Acetobacter* (25.5%), and an enteric bacteria cluster (10.4%) that is mainly comprised of Pectobacterium (4.8% of total average relative abundance), Serratia (3.5%),

Erwinia (1.3%), and Brenneria (0.5%). In contrast, *Gluconobacter* and the enteric bacteria

cluster are virtually absent from our lab-reared flies (<0.001 and <0.1%). *Acetobacter* is

extremely common in UCSD Stock Center lab-reared flies (72.7%), but comprises only 1.2% of

the bacterial communities in flies from the Petrov lab. On the other hand, *Lactobacillus*

contributes a substantial fraction of sequences in lab-reared flies (60.4% in Petrov lab, 19.1%

UCSD Stock Center) while playing only a minor role in wild-caught flies (0.5%). In addition,

*Leuconostoc* is common in the Petrov lab (28.0%) but rare (1%) in wild-caught flies and the

UCSD Stock Center (<1%). Inspection of individual samples revealed that the relative abundance of *Leuconostoc* is highly variable across *D. melanogaster* and *D. simulans.* In Petrov lab flies, relative abundance ranges from 87.6% and 84.5% in samples m.pet1 and s.pet1, respectively, to being undetectable in m.pet4, m.pet5, s.pet2, and s.pet5 (Figure 2C).

198

In addition to differences in broad patterns of community composition, we also detected two wild-caught samples dominated by genera that are rare overall: 80.3% of all sequences in the *D. melanogaster* sample m.ora1 collected from oranges were classified as *Enterococcus* (80.3%), while the sample m.str collected from strawberries has a high prevalence of *Providencia* (26.3%). The relative abundance of *Enterococcus* is smaller than 0.5% in all other wild-caught samples. *Providencia* was detected in only three other samples at a relative abundance smaller than or equal to 1%.

Intriguingly, 92% (1165 sequences) of all *Providencia* sequences from sample m.str are identical, suggesting the presence of a single, high-frequency *Providencia* strain in m.str. The highly prevalent sequence from sample m.str is 100% identical to the sequence of *P. alcalifaciens* from Juneja and Lazzaro [38], while it differs from all other *Providencia* sequences in [38] by at least two positions (Figure 3A). *P. alcalifaciens* was shown to be highly virulent in *D. melanogaster* [7] causing the highest mortality amongst all strains tested and reaching cell counts of up to $10^6$ colony forming units per fly.

By grouping all sequences into 97% identity OTUs we sought to obtain a more detailed picture of bacterial community composition. Figure 3B depicts the relative abundance of the ten most abundant OTUs across all samples. A single OTU classified as *Gluconobacter* is common among all wild-caught flies (34.7% average relative abundance, OTU 25), but completely absent from lab-reared flies. Even in the wild-caught fly sample m.ora1 that is dominated by an *Enterococcus* OTU (OTU 60) this *Gluconobacter* OTU represents 8.9% of all non-*Enterococcus* sequences. Because this OTU is common in all wild-caught flies, and specific to wild-caught flies, it is a strong candidate for being a member of the *Drosophila* core microbiome in nature. Three *Acetobacter* OTUs are also common in wild-caught flies (OTUs 26, 23, and 29). However, these OTUs are rare in flies collected from oranges and OTU 26 is also prevalent in lab-reared flies from the UCSD Stock Center. In lab-reared flies, especially flies from the Petrov lab, three *Lactobacillus* OTUs are common (OTU 28, 22, and 35). The abundance of these OTUs is highly variable between samples, with one dominant OTU (OTU 28) that is common in most Petrov lab

228  samples, while the other two OTUs are at high frequency in the larval samples mpet1_l (OTU
229  22) and m.pet6_l (OTU 35). The second most common *Acetobacter* OTU (OTU 38) is common
230  only in the UCSD Stock Center samples and larval sample s.pet3_l.  In UCSD samples, this
231  OTU is strongly negatively correlated with OTU 26  ($P = 2.9$ x $10^{-5}$ , $r^2 = 0.64$), which was also
232  classified as *Acetobacter*.

233

234  The composition of bacterial communities associated with flies differ between laboratories and
235  the wild

236  In order to further explore the factors shaping the observed variation in bacterial communities
237  between lab-reared and wild-caught flies, we carried out a Principal Coordinate Analysis (PCoA)
238  using pairwise Jaccard distances. Jaccard distances compare the number of OTUs that are shared
239  between two communities to the total number in both communities, with a smaller proportion of
240  shared OTUs leading to an increased Jaccard distance. Jaccard distance analysis requires that the
241  same number of sequences is used in each sample. This is because samples that contain more
242  sequences are more likely to include low frequency OTUs that can appear private to that sample
243  and inflate Jaccard distances. We therefore *in silico* capped the number of sequence reads per
244  sample to a common number by subsampling. In order to test for potential stochastic effects of
245  subsampling on our results, we analyzed 1000 bootstraps of the subsampling for all PCoAs
246  presented.

247

248  Figure 4A shows the position of all samples analyzed in this study relative to the first two PCos.
249  PCo1 explains 16.1% of the variation and separates wild-caught, Petrov lab, and UCSD Stock
250  Center communities from each other ($P < 5.4$ $10^{-15}$ and $r^2 = 0.79$, ANOVA, 100% of bootstraps $P$
251  $< 1.9$ x $10^{-11}$). PCo2 explains 9.9% of the variation and separates wild-caught from lab-reared
252  flies ($P < 2$ x $10^{-16}$ and $r^2 = 0.87$, ANOVA, 100% of bootstraps $P < 9.3$ x $10^{-13}$). These results
253  suggest that Petrov lab, UCSD Stock Center, and wild-caught flies all have their own distinct
254  bacterial communities.

255

256  Similarity between larval and adult samples from the same laboratory further underscores the

257　importance of the origin of the flies (Petrov lab, UCSD Stock Center) for the composition of

258　their associated microbiota. The only exception is Petrov lab *D. simulans* larval sample 3

259　(s.pet3_l), which grouped closer to the UCSD samples in Figure 4A and has a more UCSD-like

260　community dominated by *Acetobacter* (Figure 2C).

261

262　<u>Communities of wild-caught flies differ by substrate and between *D. melanogaster* and *D.*</u>

263　<u>*simulans*</u>

264　We analyzed paired samples of wild-caught *D. melanogaster* and *D. simulans* isolated from five

265　different natural substrates (oranges, apples, peaches, strawberries, and compost) in order to

266　elucidate the influence of substrate on fly-associated bacterial communities in the wild. Figure

267　4B shows a PCoA including only wild-caught *D. melanogaster* and *D. simulans* samples.

268　Communities of flies collected from oranges at three different sampling locations are clearly

269　separated from the remaining samples by PCo1 ($P = 0.00017$, $r^2 = 0.71$, ANOVA, 100% of

270　bootstraps $P < 0.008$). PCo2 separates bacterial communities from the flies collected from the

271　compost pile and those from the flies collected from the fruit substrates ($P < 0.001$, $r^2 = 0.61$,

272　ANOVA, 98.1% of bootstraps $P < 0.05$), indicating that food substrate or a variable correlated

273　with food substrate is an important factor shaping fly-associated bacterial communities.

274　Interestingly, communities of flies from strawberries, apples and peaches are relatively similar

275　irrespective of sampling location. Flies from strawberries were collected from a sampling

276　location on the West coast of the US while flies from apples and peaches were collected on the

277　East Coast of the US.

278

279　While the first two PCos in the PCoA of wild-caught flies (Figure 4B) reflect differences related

280　to food substrate, PCo3, PCo4, and PCo5 reveal a more subtle, but significant difference between

281　the communities associated with the two fly species. In 78% of all subsampling bootstraps, we

282　found a significant difference (ANOVA $P < 0.05$) between *D. melanogaster* and *D. simulans*

283　associated microbial communities along these PCos (Supplementary Figure 1). This represents a

284　significant enrichment of low p-values ($P < 4.9$ x $10^{-149}$, Chi-squared test). An example from

285　these bootstraps, in which PCo3 differentiates between *D. melanogaster* and *D. simulans,* is

286  given in Figure 4C ($P = 0.0011$, $r^2 = 0.60$, ANOVA). We do not detect such a difference between
287  lab-reared *D. melanogaster* and *D. simulans* (data not shown).

288

## Discussion

290  In this study we focused primarily on understanding the factors that shape *Drosophila*-associated
291  bacterial communities, with an emphasis on the relative roles of environmental and host species
292  effects. In order to disentangle environmental from host species effects, we collected and
293  compared sample pairs of *D. melanogaster* and *D. simulans* from the same natural substrates. We
294  extended this approach by analyzing these two species under controlled laboratory conditions.
295  Finally, in order to generalize our results, we also analyzed a set of host species spanning the
296  *Drosophila* phylogeny. A correlation between genetic distance of different fly species and the
297  dissimilarity of their bacterial communities under controlled conditions would be an indication
298  that genetic differences between host species could play a role in shaping fly bacterial
299  communities. Therefore, we extracted bacterial DNA from whole flies by carrying out extensive
300  tissue homogenization. The bacterial load on the fly surface is known to be ~10 times lower than
301  the interior load [29]. Therefore, the influence of external bacteria on the total community
302  composition is expected to be rather minor. Additionally, our focus on the total bacteria
303  associated with the whole fly, and not only the intestinal tract, was motivated by the belief that
304  bacteria associated with fly surfaces might play important roles in shaping the fly environment.
305  This is supported by Ren *et al*. [29] who found acetic acid bacteria accumulating in bristled areas
306  on the fly surface, likely forming biofilms and by Barata *et al*. [33] who demonstrated that
307  damaged grapes do not acquire acetic acid bacteria when insects, particularly *Drosophila,* are
308  physically excluded. We therefore do not expect the   These acetic acid bacteria could very well
309  be transported on the fly surface. Note that even though we aspirated flies from individual fruit
310  and attempted to associate bacterial communities with the substrate, we likely sampled bacterial
311  communities that the fly has acquired during its life span. This includes bacteria from the
312  particular fruit from which it was sampled, but could also include bacteria potentially from prior
313  locations.

<u>Factors shaping natural fly associated communities</u>

We determined that substrate or a strongly correlated variable is the most important factor shaping bacterial communities in wild-caught flies. Diet has been previously suggested as a major determinant of bacterial community composition in mammals [14,15,39] and flies [32] and our results agree with these findings. The most distinct bacterial communities were associated with flies collected from oranges. Oranges contain citric acid and might have a lower pH than other substrates. Furthermore, orange peel contains essential oils that have bactericidal properties that might influence the bacterial community composition [40]. Although the substrate appears to be a plausible factor shaping the communities here, we cannot disentangle its effects from seasonal effects (e.g. temperature, humidity). This is because we collected flies from different substrates at different times of the year when the respective fruit were ripe.

We carefully sampled *D. melanogaster* and *D. simulans* across different sites and substrates in nature which allowed us to disentangle environmental effects from host species effects on microbial community composition. We found evidence that host fly species identity (*D. melanogaster* vs. *D. simulans*) detectably influences the associated microbial communities, but that the effect is subtle. Although our power comparing lab-reared *D. melanogaster* and *D. simulans* might be lower because of smaller sample size and restriction to fewer sequences, mainly due to high *Wolbachia* prevalence in some lab-reared samples, it is intriguing that this host species effect is detectable only in the wild and could not be detected in lab-reared flies. Moreover, while we detected differences between two closely related sister species in the wild, we could not detect any differences for nine substantially more divergent *Drosophila* species in the lab. We found no correspondence of distances between bacterial communities and genetic distances between nine lab-reared fly species, unlike Ochman *et al.* [16] and Ley et al. [15], who found this correlation in primates and other mammals. Taken together these findings imply that the effects of host species on microbial communities are rather subtle in drosophilids and/or need natural environmental conditions to manifest themselves.

The observed difference between *D. melanogaster* and *D. simulans* microbial communities might be caused by a variety of host-associated factors, such as arrival times at fruit [41,42], age distributions in the wild (Emily Behrman and Paul Schmidt, University of Pennsylvania, personal communication), or host genetic differences [12,25,43].

### Composition of bacterial communities in the lab and in the wild

PCoA revealed that, in concordance with earlier studies [30,32], bacterial communities associated with *Drosophila* differ sharply between different laboratories and between laboratories and the wild. Interestingly, bacteria from different genera, but with similar metabolic properties, dominate the communities of wild-caught, Petrov lab, and UCSD Stock Center flies. *Gluconobacter* species are the most prevalent bacteria in wild-caught flies in our study. This is in accordance with Corby-Harris *et al*. [31], who also find abundant *Gluconobacter* sequences in wild-caught flies, but different from Chandler et al. [32] who find a smaller fraction of *Gluconobacter* sequences. More than 90% of all *Gluconobacter* sequences in our study can be grouped into a single OTU that is common in all wild-caught flies. In contrast, *Gluconobacter* is almost absent from the lab strains. Thus, this OTU is a strong candidate for being a major member of a core microbiome that is shared among and specific to wild-caught *D. melanogaster* and *D. simulans*. *Gluconobacter* belongs to the same family (Acetobacteraceae) as *Acetobacter*, which is also common in wild-caught flies with the exception of flies from oranges that carry less *Acetobacter*. *Acetobacter* is also the most prevalent genus in flies from the UCSD Stock Center and has very similar metabolic capabilities. Both genera, *Gluconobacter* and *Acetobacter*, oxidize sugars and alcohol to acetic acid, and tolerate low pH as well as high ethanol concentrations [44]. Acetic acid bacteria have been reported to occur in association with many insect species and a role as important symbionts has been postulated by Crotti et al. [45]. Lactobacilli, which are at high prevalence in Petrov lab flies, tolerate low pH and high ethanol concentrations as well, but instead oxidize sugars to lactic acid [46]. The high prevalence of bacteria with similar metabolic capabilities, tolerance of low pH, and high ethanol concentrations strongly suggests that there is environmental selection for these bacterial groups. Rotting fruit, the most important natural substrate for *D. melanogaster* and *D. simulans* in our study, contain high amounts of sugar and are known to be colonized by a variety of ethanol producing yeasts [47]. Yeasts can produce high alcohol concentrations, thereby generating a nutrient rich environment for acetic acid or lactic acid producing bacteria (Acetobacteraceae and Lactobacillaceae), while inhibiting the growth of those less tolerant to alcohol. The production of these acids selects for acid tolerant microorganisms including the microorganisms that produced the acids in the first place. This suggests that environmental selection [48] is an important factor for the observed prevalence of these bacteria.

Interestingly bacteria of the genus *Lactobacillus*, which have been associated with effects on *Drosophila* growth [28] and even assortative mating [49], are prevalent only in the lab in our study. Sixty percent of all sequences from Petrov lab flies, and 19% of all sequences obtained from UCSD Stock Center flies are *Lactobacillus*. In most wild-caught samples *Lactobacillus* represented less than 1% of all sequences. This finding is corroborated by results from Chandler *et al.* [32], who find an increase of the proportion of *Lactobacillus* species in lab-reared flies. Thus, while studying the effects of *Lactobacillus* on drosophilids in the laboratory is useful as a general model for insect-microbe interactions, its relevance to *Drosophila* in nature may be limited.

In contrast to Chandler *et al.* [32], who found that Enterobacteriaceae from group Orbus are highly prevalent in *Drosophila*, these bacteria are absent or at very low frequency in our samples (not amongst the best BLAST hits for any of the 100 most abundant OTUs in our data set). We can only speculate about the reasons for this difference here. One possibility might be an epidemic of Orbus group bacteria in 2007 and 2008, when Chandler *et al.* [32] collected their samples.

Given the strong effect of food substrates that we observed in wild *Drosophila*, similar effects might play a role in lab-reared flies. Differences in the provided food substrates between laboratories might therefore lead to differences in communities. For example, we provide our flies with a corn meal molasses diet, whereas the Stock Center uses sugar instead of molasses. In addition, our food contains Tegosept(r) to reduce microbial growth, while this ingredient is only optional at UCSD. Intriguingly, Chandler *et al.* [32] found that fly-associated bacterial communities differed between labs at UC Davis despite using the same food from the same kitchen, suggesting that other factors are involved as well. Candidate explanations would involve ecological drift, which is likely to be stronger in the laboratory, and priority effects [42,50,51]. A potential role of stochastic drift processes and priority effects is supported by the notion that the occurrence of the two major Acetobacter OTUs (OTU 26 and 38) in the UCSD Stock Center flies is strongly antagonistic. This is in accordance with a model in which one of the OTUs quickly occupies an ecological niche and excludes its ecologically similar, close relative.

Bacterial communities of lab-reared flies are highly variable in diversity and composition within and between laboratories in this study. Because fly phenotypes are influenced by bacteria [27,28,52], this bacterial variation can add to the variance of phenotypic traits. This makes it more difficult to detect genetic variation underlying phenotypic traits and reduces reproducibility between laboratories. The presence of a certain microbiota might also lead to unwanted results in genetic trait mapping: Genetic variation that is attributed to directly underlie a phenotypic trait might indeed interact with microbes that influence this trait instead, thus influencing the trait only indirectly. Monitoring of microbial communities during experiments in which phenotypes are measured could be a means to approach these difficulties.

Species richness of lab-reared and wild-caught *Drosophila* associated bacterial communities
Although diversity varies strongly across different samples from lab-reared flies, their bacterial communities are on average less diverse than those of wild-caught. This has been reported previously [30,32,53].

The three most plausible explanations for this pattern in our study are: (i) laboratory fly food is highly homogeneous and contains antimicrobial preservatives, proprionic acid and Tegosept(r) in our case, which inhibit bacterial growth and likely reduce bacterial diversity, (ii) the transfer of

429 flies to vials with fresh food during stock keeping could lead to ecological drift [50], which
430 reduces reduces diversity in the long run due to potential loss of taxa, (iii) while there is a
431 constant influx of new bacteria into natural fly habitats, e.g. from other insects or via aerial
432 transport, this influx is limited by cotton-sealed vials used in *Drosophila* husbandry.
433
434 It is known that species richness is often overestimated using pyrosequencing approaches (e.g.
435 [54]). We applied rigorous quality filtering and Chimera detection (see Materials and Methods)
436 and used an OTU threshold of 97% identity which is thought to be robust against sequencing and
437 PCR errors [54]. Although we take all these measures, we can not exclude that we are still
438 overestimating the diversity in our samples. On the other hand overly stringent removal of
439 sequences might make us miss important aspects of microbial communities [55].
440
441 <u>Potential fly pathogens</u>
442 The bacterial communities of certain wild-caught fly isolates contained potential *Drosophila*
443 pathogens at high frequencies. In one sample of *D. melanogaster* from strawberries, more than
444 25% of all sequences were identical to those of *P. alcalifaciens* whereas *Providencia* is absent or
445 at very low frequency in all other samples. This bacterium is known to be highly virulent in fruit
446 flies [7], but reaches high bacterial loads in flies usually only when flies are systemically infected
447 (personal communication, Brian Lazzaro, Cornell University). *Enterococcus* was present at high
448 abundance in one *D. melanogaster* orange sample 1 (m.ora1, 80.3%), but virtually absent from
449 all other samples. *Enterococcus* species were previously found to be associated with *D.*
450 *melanogaster* [32] and are highly prevalent in the lab-reared flies studied by Cox and Gilmore
451 [30]. These authors showed that *Enterococcus* can reach densities of $10^5$ colony forming units
452 per fly, causing severe disease symptoms and high mortality. This compares to a total of ~$10^4$
453 colony forming units including all bacterial species in healthy flies [29,30].
454
455 The presence of these disease-associated genera in individual samples, and their absence or near
456 absence from other samples suggests that one or more flies were systemically infected in the
457 samples that showed a high relative abundance of the disease associated genus . Thus, detection

458   of infections with potential pathogens in natural fly populations seems possible by bacterial 16S

459   rRNA gene sequencing. Hence, 16S rRNA sequencing could be a powerful means for the

460   epidemiological monitoring of bacterial pathogens.

461

**462 Conclusion**

463   We show that under natural conditions the bacterial communities associated with *Drosophila*

464   correlate mainly with the substrate the flies have been collected from and to a smaller extent with

465   fly species. Despite appreciable effort, we did not find evidence for host species effects on the

466   bacterial communities under controlled laboratory conditions. Instead, laboratory of origin and

467   stochastic effects on microbial communities are pronounced in the laboratory. This suggests that

468   host genetic effects, as represented by genetic differences between the fly species in this study,

469   might be rather small or absent in the lab, while there is potential for such effects under natural

470   conditions. Furthermore, we find that acetic acid producing bacteria (Acetobacteracea) are

471   ubiquitous symbionts of *Drosophila* in nature. Intriguingly, it has been shown both that *D.*

472   *melanogaster* promotes dispersal and establishment of these bacteria [33] and that the presence

473   of acetic acid bacteria can have beneficial effects on *D. melanogaster* larval growth and

474   development time [27]. Together these findings suggest that *D. melanogaster* and its siblings

475   transport and establish the acetic acid bacteria on the substrates, which might modify these

476   substrates in ways beneficial to the flies and their offspring. We speculate that the microbial

477   community associated with *Drosophila* can be seen as an external organ of the fly holobiont [56]

478   in a similar way that the human gut flora has been referred to as the "forgotten organ" [57].

479

**480 Materials and Methods**

481   <u>Fly samples</u>

482   *D. sechellia* (4021 0248.27), *D. erecta* (14021 0224.00), *D. yakuba* (14021 0261.01), *D.*

483   *persimilis* (14011-0111.49), *D. pseudoobsura* (14011-0121.148), *D. mojavensis* (15081-1351.30)

484   and *D. virilis* (15010-1051.00) were obtained from the UCSD Stock Center as well as one

485   additional *D. melanogaster* (14021-0231.131) and one *D. simulans* (14021-0251.250 ) strain.

486   The UCSD Stock Center strain ID numbers are in parentheses.

487 Petrov lab *D. melanogaster* and *D. simulans* were originally collected in Portland, OR and San
488 Diego, CA in 2008 and lab-reared on standard molasses corn meal diet for ~3 years (27g Agar,
489 75g corn meal, 200ml molasses, 42g dry active yeast, 40ml Tegosept, 15ml propionic acid, in
490 2.8l deionized water). Note that the food is boiled for 20 minutes killing most of the microbes in
491 the food and that Tegosept is added after cooling down to prevent excessive microbial growth.
492 We used flies from six independently acquired isofemale lines from each fly species (m.pet1-6
493 and s.pet 1-6). All lines were kept under the same conditions and on the same food, but in
494 independent vials. DNA extractions and library preparations, were performed independently for
495 each line.
496
497 All adult lab-reared flies were transferred to fresh Petrov lab food vials 24 hours prior to DNA
498 extraction. Petrov lab flies were taken from culture vials in the Petrov lab and placed on fresh
499 food 24 hours prior extraction. UCSD Stock Center flies were taken from the vials we received
500 from the Stock Center and placed on fresh Petrov lab food 24 hours prior extraction.
501
502 Wild *D. melanogaster* and *D. simulans* from rotting apples, peaches, and a compost pile were
503 collected in an orchard on the East coast of the USA (Johnston, RI) in August 2010. Flies from
504 oranges were collected from three locations in the Central Valley of California USA: at a location
505 close to Brentwood, at a site East of Manteca, and a site in Escalon in February and March 2011.
506 Flies from strawberries were collected close to Waterford, CA in May 2011. All sampling sites
507 were at least 10km apart from each other. In most cases pairs of *D. melanogaster* and *D.*
508 *simulans* were picked from the same individual fruit. Otherwise, flies were selected from the
509 same type of fruit in close proximity. Flies were transported to the lab alive, in empty vials. On
510 hot days, flies were slightly chilled using ice or car A/C. All flies were brought back to the lab
511 within 5 hours of collection. Males of *D. melanogaster* and *D. simulans* were identified by
512 genital morphology and stored at -80°C until DNA extraction. Flies from Johnston, RI were
513 shipped on dry ice to the Petrov lab for DNA extraction.
514
515 For the collection of larval samples from lab-reared flies, adult flies were transferred to fresh

516  Petrov lab food vials for two days and then removed from the vial again. Vials containing eggs
517  were kept at room temperature until larvae started to crawl out of the food for pupation. Larvae
518  leaving the food and larvae of the same size that were still in the food were regarded third instar
519  larvae and collected for DNA extraction. Excess food was removed from the larvae by
520  transferring them to a microcentrifuge tube containing 500μl PBS (pH 7.4), vortexing for 3
521  seconds, and then discarding the liquid. The larval samples correspond to the adult flies i.e. the
522  sample named m.pet1_l was collected from the same isofemale line as m.pet1 using the
523  procedure described above.
524
525  <u>DNA Extraction and PCR</u>
526  DNA was extracted from pools of five males, with the exception of *D. simulans* orange sample 1
527  (s.ora1) and *D. melanogaster* orange sample 3 (m.ora3), for both of which we were able to
528  retrieve three males only. Larval samples included three third instar larvae per sample. DNA
529  extraction was performed using the Qiagen QIAamp DNA extraction kit (Qiagen, Carlsbad,  CA)
530  following the manufacturer's protocol with the following modifications: Flies/larvae were
531  incubated in buffer ATL containing proteinase K at 56°C for 30 min to soften and predigest the
532  exoskeleton. Digestion was then interrupted by 3 minutes of bead beating on a BioSpec Mini
533  Bead Beater 96 with glass beads 0.1mm, 0.5mm, and 1mm in size (BioSpec, Bartlesville, OK),
534  followed by another 30 min of incubation at 56°C. After addition of lysis buffer AL samples
535  were incubated 30min at 70°C and 10min at 95°C. The remaining extraction procedure was
536  performed according to the manufacturer's protocol. Extraction controls were run in parallel with
537  all samples to monitor contamination. Broad range primers (27F and 338R) were fused to
538  identification tags and the 454 sequencing primers to amplify a fragment spanning the variable
539  regions V1 and V2 of the bacterial ribosomal 16S rRNA gene. The primer sequences are (5´-
540  CTATGCGCCTTGCCAGCCCGCTCAGTC<u>AGAGTTTGATCCTGGCTCAG</u>-3´) and reverse (5
541  ´-CGTATCGCCTCCCTCGCGCCATCAGXXXXXXXXXXXCA<u>TGCTGCCTCCCGTAGGAGT</u>-
542  3´).  The Xs are a placeholder for identification tags (Multiplex Identifiers, MIDs); a different tag
543  was used for each amplification reaction. Primers 27F and 338R are underlined. DNA was
544  amplified using Phusion® Hot Start DNA Polymerase (Finnzymes, Espoo, Finland) and the

545  following cycling conditions: 30 sec at 98°C; 35 cycles of 9 sec at 98°C, 30 sec at 55°C, and 30

546  sec at 72°C; final extension for 10 min at 72°C). In order to reduce PCR bias, amplification

547  reactions were performed in duplicate and pooled. In order to reduce the number of *Wolbachia*

548  amplicons, PCR products were restriction digested with 2μl FastDigest® BstZ17 (Fermentas,

549  Glen Burnie, MD) at 37°C for 30 min. BstZ17 was selected to specifically cut *Wolbachia*

550  sequences close to the middle of the amplified region. Reaction products were run on an agarose

551  gel, extracted using the Qiagen MinElute Gel Extraction Kit and quantified with the Quant-iT™

552  dsDNA BR Assay Kit on a NanoDrop 3300 Fluorometer. Equimolar amounts of purified PCR

553  product from each sample were pooled and further purified using Ampure Beads (Agencourt).

554  The pool was run on an Agilent Bioanalyzer prior to emulsion PCR for final quantification.

555  Resulting PCR products were run on a 454 sequencer using Titanium Chemistry. A set of

556  samples was extracted using a FastPrep FP120 bead beater (Qbiogene, Carlsbad, CA). These

557  samples include *D. erecta*, *D. yakuba*, *D. sechellia* from the UCSD Stock Center, and wild-

558  caught samples collected in Johnston, RI. These samples were sequenced twice, with and without

559  the BstZ17 digest. Relative abundance of bacterial taxa correlated strongly between the two

560  procedures for these samples after removal of *Wolbachia* reads (mean $r^2 = 0.94$). Therefore, we

561  pooled the sequencing reads obtained with and without the digest to get a higher sequencing

562  depth per sample.

563

564  Amplicons from samples with a high *Wolbachia* load were often so effectively digested that the

565  final DNA yield was too small for library preparation. In order to have enough PCR-product for

566  library construction, we shortened digestion time for amplicons from these samples to 5 minutes,

567  resulting in an incomplete digest. Predictably, these samples yielded a high percentage of

568  *Wolbachia* sequences after the incomplete digest.

569

570  We verified the specificity of the BstZ17 for cutting *Wolbachia* sequences by an *in silico* search

571  for restriction sites in our sequences from undigested samples, all sequences from Chandler *et al.*

572  [32], and all bacterial sequences in the SILVA data base. A very small fraction of non-*Wolbachia*

573  sequences would have been cut in our sequence set from undigested samples (27 out of 23423)

20  

574 and the data from Chandler *et al.* [32] (18 out of 3243, mainly confined to a single sample). The

575 majority of these sequences were classified as Rhizobiales. *In silico* search for the BstZ17

576 restriction site in sequences from the SILVA database revealed that the sequences that would

577 have been cut by the restriction enzyme fall mainly into the orders of Rhizobiales,

578 Myxococcales, and a non-*Wolbachia* Rickettsiales. Although these orders have either not been

579 reported to be associated with *Drosophila* or occur only at very low numbers, the pretreatment

580 with BstZ17 of most of our samples might have led to underestimation of their abundance in this

581 study.

582

583 <u>Data analysis</u>

584 The MOTHUR v1.23.1 [37] software was used for analysis. We used the trim.seqs command to

585 remove primer and MID tags and quality filter our sequences according to the following

586 requirements: Minimum average quality of 35 in each 50 bp window, minimum length of 260 bp,

587 homopolymers no longer than 8 bp. Only sequences matching the MIDs and the bacterial

588 primers perfectly were kept. Passing sequences were filtered for sequencing errors using the

589 pre.cluster command. Sequences were then screened for chimeras using UCHIME [58] as

590 implemented in MOTHUR with standard settings separately for each sample. 2% of all

591 sequences were identified as chimeric and discarded. The remaining sequences were aligned to

592 the SILVA reference database [36] using the MOTHUR implemented kmer algorithm with

593 standard settings. Sequences not aligning in the expected region were removed using the

594 screen.seqs command. Sequences were classified into bacterial taxa with the classify.seqs

595 command using the SILVA reference database and taxonomy with default settings. Sequences

596 classified as *Wolbachia* were removed from further analysis. Grouping of sequences into OTUs

597 was done using the MOTHUR implemented average neighbor algorithm. Inverted Simpson and

598 Shannon diversity indices were generated with the collect.single command. Rarefaction sampling

599 was performed with the rarefaction.single command. The sequence with the smallest distance to

600 all other sequences in each  OTU was picked with the get.oturep command using the weighted

601 option and classified with the classify.otu command using the SILVA reference database and taxonomy. Representative sequences of the 100 most common OTUs were also searched in the nr/nt database of the National Center for Biotechnology Information (ncbi) using megablast with default settings via the web server (http://blast.ncbi.nlm.nih.gov/). Taxonomy information from the BLAST results was compared to the classification using the SILVA database. PCoA of Jaccard distances was performed applying the pcoa command on a Jaccard distance matrix generated with the dist.shared command. Because Jaccard distance is based on presence and absence of OTUs, it is sensitive to information from low abundance OTUs, even in the presence of other more abundant OTUs. In an abundance-based distance measure this information would likely be swamped by few extremely common OTUs. These considerations are particularly relevant to our study where a handful of bacterial families dominate the data (Figure 2A). Jaccard distances are also less prone to be affected by biased abundance measurements that can result from amplification biases during PCR amplification of the bacterial 16S rRNA gene. The downside of the sensitivity of Jaccard distances to low abundance OTUs is that samples with a higher number of bacterial sequence reads can be biased towards detecting more low abundance OTUs which inflates Jaccard distance. Therefore the number of sequences per sample was *in silico* capped to have the same number of sequences per sample before calculation of Jaccard distances. The caps were  912 sequences per sample for the PCoA of wild-caught flies and 116 sequences per sample for the PCoA including all samples.

**Acknowledgements**

We thank Alan Bergland for providing samples and for helpful comments on the manuscript. We thank Silke Carstensen and Heinke Buhtz for excellent technical assistance. We thank Angus Chandler, Artyom Kopp, Jonathan Eisen and Brian Lazzaro for helpful discussions; and Anna Sophie Fiston-Lavier, Diamantis Sellis, Heather Machado, David Enard, and Nandita Garud for helpful comments on the manuscript. We thank the UCSD Stock Center for providing samples.

sensitivity and speed of chimera detection. Bioinformatics 27: 2194–2200. doi:10.1093/bioinformatics/btr381.

630

631

632

633

634

635

636

637

638

639

640

641

642

643

644

645

646

647

648

649

650

651

652

653

654

655

656

**Tables**

658 Table 1 Sample list.

| sample name | species | substrate | location | n | larva |
|---|---|---|---|---|---|
| m. | D. melanogaster | lab diet | UCSD Stock Center | 1 | yes |
| s. | D. simulans | lab diet | UCSD Stock Center | 1 | yes |
| sech | D. sechellia | lab diet | UCSD Stock Center | 1 | yes |
| yak | D. yakuba | lab diet | UCSD Stock Center | 1 | yes |
| erec | D. erecta | lab diet | UCSD Stock Center | 1 | yes |
| pers | D. persimilis | lab diet | UCSD Stock Center | 1 | yes |
| pseu | D. pseudoobscura | lab diet | UCSD Stock Center | 1 | yes |
| vir | D. virilis | lab diet | UCSD Stock Center | 1 | yes |
| moja | D. mojavensis | lab diet | UCSD Stock Center | 1 | yes |
| m.pet1-m.pet6 | D. melanogaster | lab diet | Petrov lab | 6 | yes |
| s.pet1-s.pet6 | D. simulans | lab diet | Petrov lab | 6 | yes |
| m.app | D. melanogaster | apple | Johnston, RI | 1 | no |
| s.app | D. simulans | apple | Johnston, RI | 1 | no |
| m.pea | D. melanogaster | peach | Johnston, RI | 1 | no |
| s.pea | D. simulans | peach | Johnston, RI | 1 | no |
| m.com | D. melanogaster | compost | Johnston, RI | 1 | no |
| s.com | D. simulans | compost | Johnston, RI | 1 | no |
| m.ora1 | D. melanogaster | orange | Central Valley 1/Manteca | 1 | no |
| s.ora1 | D. simulans | orange | Central Valley 1/Manteca | 1 | no |
| m.ora2 | D. melanogaster | orange | Central Valley 2/Escalon | 1 | no |
| s.ora2 | D. simulans | orange | Central Valley 2/Escalon | 1 | no |
| m.ora3 | D. melanogaster | orange | Central Valley 3/Brentwood | 1 | no |
| s.ora3 | D. simulans | orange | Central Valley 3/Brentwood | 1 | no |
| m.str | D. melanogaster | strawberry | Central Valley 4/Waterford | 1 | no |
| s.str | D. simulans | strawberry | Central Valley 4/Waterford | 1 | no |

659 n = number of samples, each sample consisting of 5 male flies with the exception of s.ora1 and

660 m.ora3 where only 3 males were available

661

662

663

664

Table 2 Comparison of bacterial community diversity with previous studies on *Drosophila*.

| study | fly type | tissue | Chao's species richness estimate | SD | Shannon diversity index | SD | Mean no. of sequences per sample | sequence type | primer position |
|---|---|---|---|---|---|---|---|---|---|
| This study | wild-caught *D. melanogaster* and *D. simulans* | whole flies | 181 (22) [43] | 97.1 (14.5) [15.0] | 1.79 (1.65) [1.63] | 0.44 (0.38) [0.34] | 3800 (100) | 454 pyro | 27F 338R |
| This study | 9 lab-reared *Drosophila* species | whole flies | 94* (15) [21***] | 91.8 (24.6) [15.9] | 0.77*** (0.69***) [0.64***] | 0.67 (0.61) [0.58] | 3580 | 455 pyro | 27F 338R |
| Wong et al. 2011 | lab-reared 3-7 day old males *D. melanogaster* | gut | 19*** [**] | n.a. | 1.26** [*] | n.a. | 113614 (single sample) | 454 pyro | 27F 338R |
| Wong et al. 2011 | lab-reared 3-5 week old males *D. melanogaster* | gut | 17*** [**] | n.a. | 0.72** [**] | n.a. | 85095 (single sample) | 454 pyro | 27F 338R |
| Chandler et al. 2011 | wild-caught *D. melanogaster* | surface washed whole flies/gut | 19*** (n.s.) | 11.5 | 2.03 n.a. (***) | 0.52 | 83 | clone library | 27F 1492R |
| Corby-Harris et al. 2007 | wild-caught *D. melanogaster* | surface washed whole flies | 24*** (n.s.) | 15.8 | n.a. | n.a. | 66 | clone library | 27F 1522R |
| Cox and Gilmore 2007 | wild-caught *D. melanogaster* | surface washed whole flies | 25*** (n.s.) | n.a. | 2.3*** (***) | n.a. | 211 | clone library | 27F 1492R |

***$P < 0.001$, ** $P < 0.01$, * $P < 0.05$ Student's T-test; n.s. = non significant; n.a. = not available. P-values are relative to wild-caught flies in this study. Values and p-values in parentheses are for subsampling our samples to 100 16S rRNA gene sequences per sample to make the results comparable to other studies on wild-caught *Drosophila*. Values in square brackets are for removing all OTUs that contain fewer than 10 sequences from the analysis to make our study more comparable to Wong *et al.* [20].

**Figures**

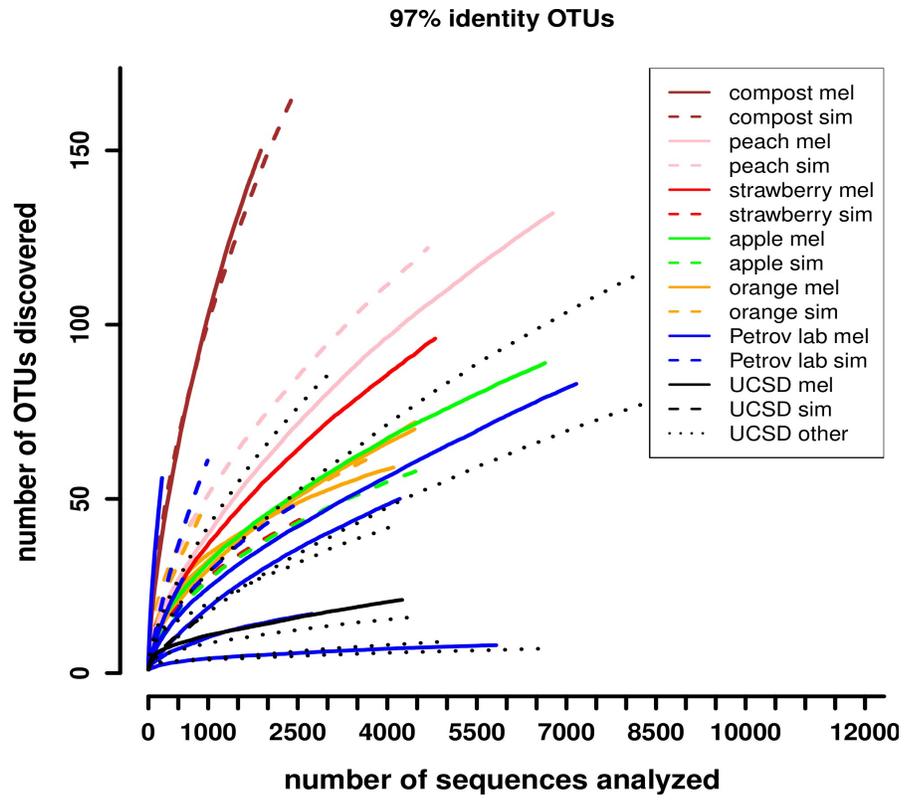

97% identity OTUs

number of OTUs discovered

number of sequences analyzed

| | |
|---|---|
| compost mel | |
| compost sim | |
| peach mel | |
| peach sim | |
| strawberry mel | |
| strawberry sim | |
| apple mel | |
| apple sim | |
| orange mel | |
| orange sim | |
| Petrov lab mel | |
| Petrov lab sim | |
| UCSD mel | |
| UCSD sim | |
| UCSD other | |

 Figure 1

 Rarefaction curves of 97% identity OTUs (A) for adult male flies.

A

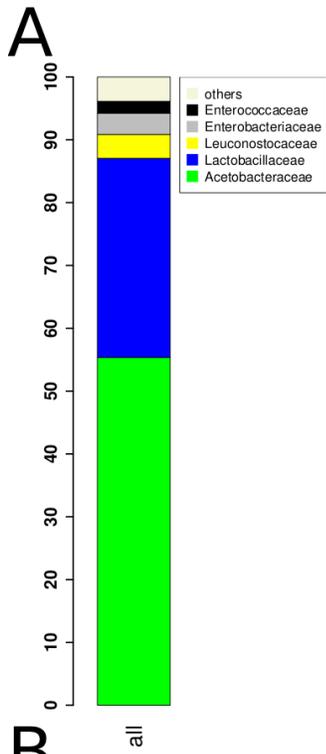

B

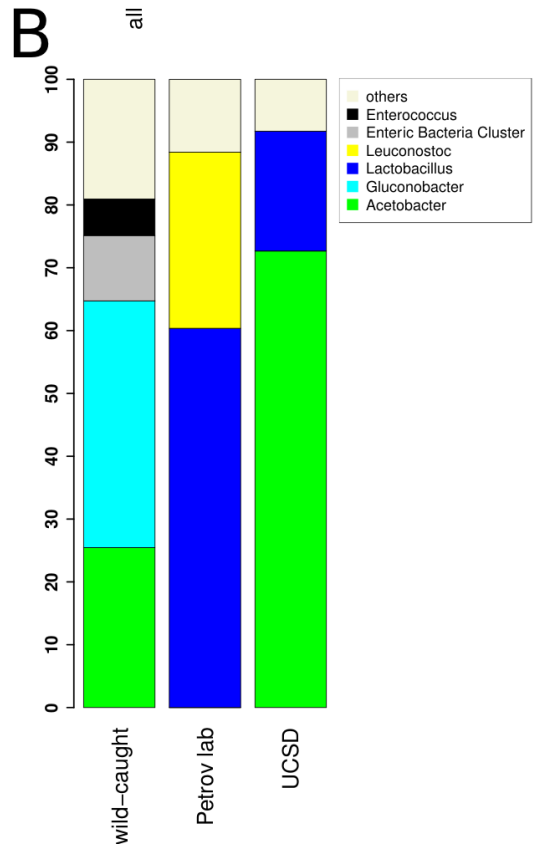

C

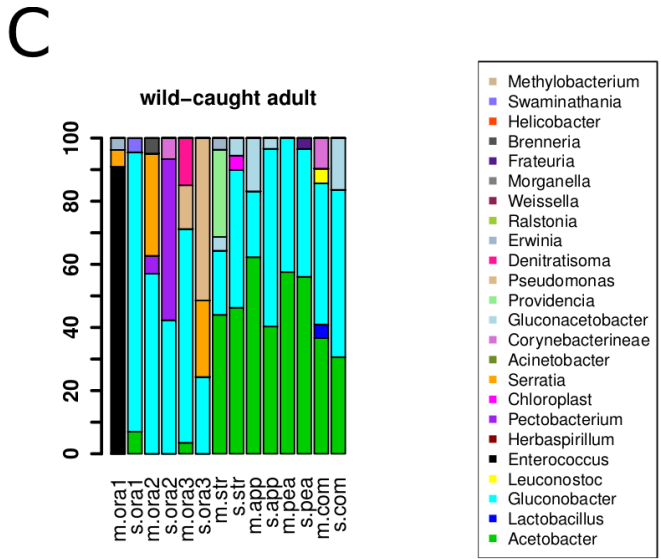

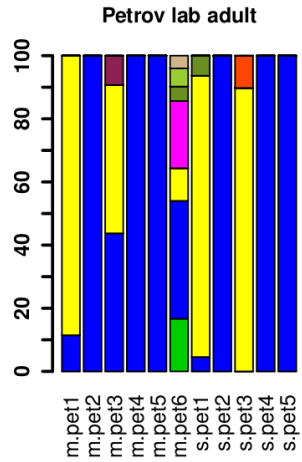

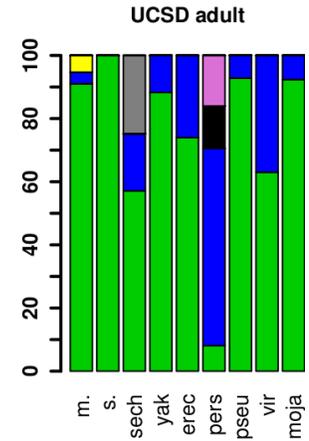

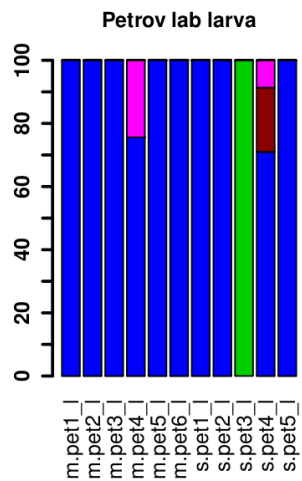

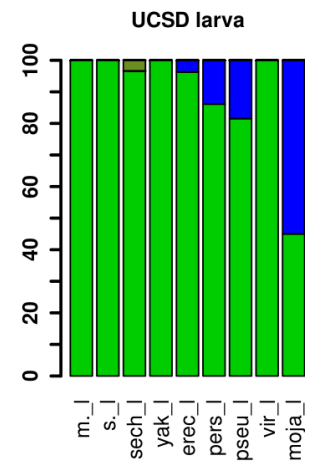

708    Figure 2

709    Relative abundance of bacterial taxa as assessed by 16S rRNA gene sequences. *Wolbachia*

710    sequences were excluded. (A) The five most abundant bacterial families associated with

711    *Drosophila* across all samples in the study. (B) Relative abundance of bacterial genera. Genera

712    present at levels less than 5% were grouped into "others" category . (C) Relative abundance of

713    bacterial genera for individual samples. Each vertical bar represents one sample of five pooled

714    male flies. Bacterial genera of abundance < 3% have been removed for clarity. *D. melanogaster*

715    sample names start with m., *D. simulans* with s.. In wild-caught samples the sample names

716    include an abbreviation for the substrate they were collected from: ora = orange, str = strawberry,

717    app = apple, pea = peach, com = compost. Names of flies from the Petrov lab contain "pet"

718    instead. Samples names ending with "_l" mark larval samples.

719

A

```
                                 1111111111
                                 670012366788
                                 690185559015
Providencia_from_m.str           TAAAATATGACA
P.alcalifaciens_DSM30120         ............
P.sneebia_DSM19967               CG.....A...T
P.rettgeri_C                     ..C....A...T
P.burhodogranariea_D             .......A...T
P.burhodogranariea_DSM19968      .G.....A...T
```

virulence

B

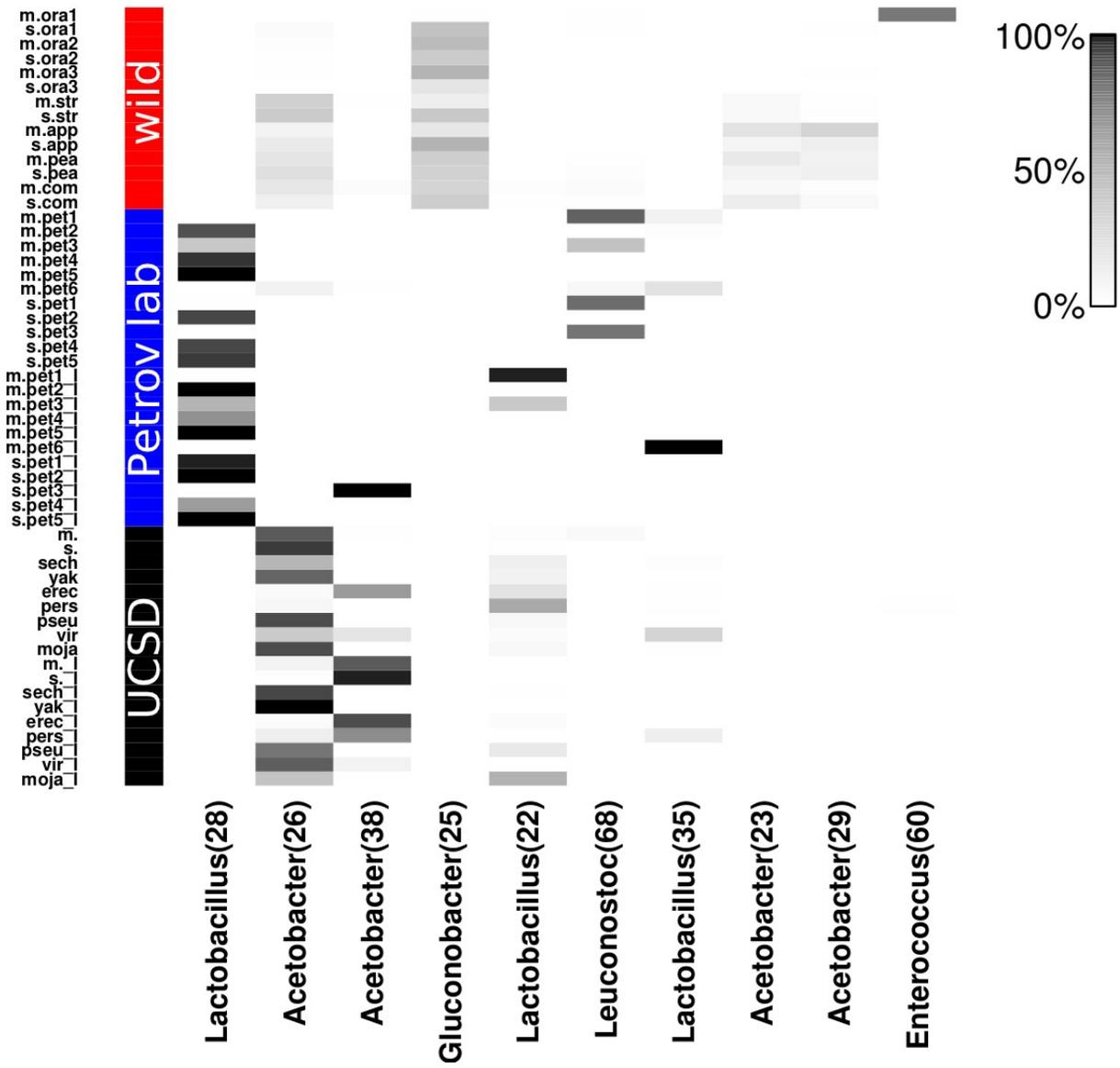

721 Figure 3

722 (A) Segregating sites of the 16S rRNA gene alignment of the highly abundant *Providencia*

723 sequence from *D. melanogaster* (grey background) collected from strawberries (m.str) to

724 *Providencia* species from [38]. Sequences are sorted by virulence as determined by [7]. Note that

725 Galac and Lazzaro determined virulence of a different but closely related *P. alcalifaciens* strain.

726 (B) Heatmap of the 10 most abundant 97% identity OTUs across all samples. OTUs are sorted

727 by average relative abundance across all samples from left to right with the most abundant OTU

728 to the left. Grey shades indicate the relative abundance of each OTU for a given sample.

729 Numbers in brackets are OTU identifiers.

730

731

732

733

734

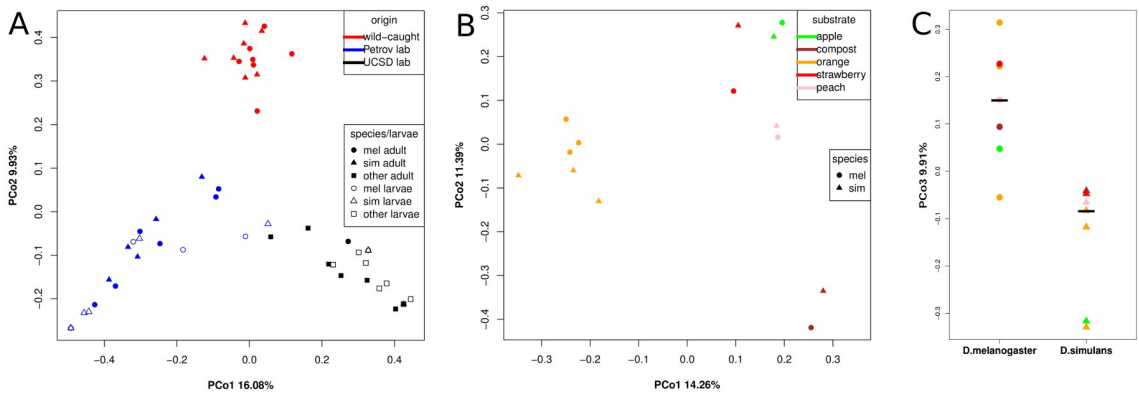

736 Figure 4

737 PCoA of Jaccard distances based on 97% identity OTUs. (A) All samples in this study. Colors

738 are according to origin. (B) Wild-caught samples. Colors are according to food-substrate (C)

739 Wild-caught samples PCo3. *D. melanogaster* and *D. simulans* differ significantly for PCo3 (*P* =

740 0.0011). Colors are according to food-substrate.

741    Supplementary Table 1 Number of sequences per sample after quality filtering before Wolbachia

742    removal.

| sample ID | sample type | number of sequences | % Wolbachia sequences |
|---|---|---|---|
| m.ora1 | adult | 4487 | 0.51 |
| s.ora1 | adult | 8944 | 89.8 |
| m.ora2 | adult | 4104 | 0.05 |
| s.ora2 | adult | 4148 | 6.65 |
| m.ora3 | adult | 4456 | 0 |
| s.ora3 | adult | 5463 | 83.23 |
| m.str | adult | 4802 | 0 |
| s.str | adult | 5395 | 53.09 |
| m.app | adult | 7366 | 9.84 |
| s.app | adult | 6912 | 33.03 |
| m.pea | adult | 6815 | 0.63 |
| s.pea | adult | 8242 | 43.24 |
| m.com | adult | 7165 | 73.72 |
| s.com | adult | 7679 | 67.99 |
| m.pet1 | adult | 2738 | 0 |
| m.pet2 | adult | 12143 | 95.25 |
| m.pet3 | adult | 7165 | 0 |
| m.pet4 | adult | 5306 | 20.67 |
| m.pet5 | adult | 5824 | 0 |
| m.pet6 | adult | 3939 | 94.19 |
| s.pet1 | adult | 11037 | 98.95 |
| s.pet2 | adult | 6618 | 98.11 |
| s.pet3 | adult | 7738 | 87.18 |
| s.pet4 | adult | 2641 | 0 |
| s.pet5 | adult | 10309 | 97.47 |
| s.pet6 | adult | 10604 | 99.83 |
| m.pet1_l | larva | 4270 | 0 |
| m.pet2_l | larva | 2329 | 5.97 |
| m.pet3_l | larva | 1880 | 0 |
| m.pet4_l | larva | 4206 | 3 |
| m.pet5_l | larva | 11846 | 0 |
| m.pet6_l | larva | 7949 | 0.74 |
| s.pet1_l | larva | 4434 | 93.82 |
| s.pet2_l | larva | 6793 | 6.07 |
| s.pet3_l | larva | 5815 | 1.44 |
| s.pet4_l | larva | 927 | 0 |
| s.pet5_l | larva | 721 | 17.06 |
| s.pet6_l | larva | 2371 | 96.88 |
| mel | adult | 8261 | 48.58 |
| sim | adult | 11048 | 92.65 |
| sech | adult | 8864 | 65.81 |
| yak | adult | 10371 | 0 |
| erec | adult | 8286 | 0.02 |
| pers | adult | 4066 | 0.02 |
| pseu | adult | 4887 | 0 |
| vir | adult | 4418 | 0 |
| moja | adult | 6590 | 0 |
| mel_l | larva | 6463 | 7.64 |
| sim_l | larva | 6294 | 4.27 |
| sech_l | larva | 3355 | 15.83 |
| yak_l | larva | 5492 | 0 |
| erec_l | larva | 4548 | 0 |
| pers_l | larva | 4614 | 0 |
| pseud_l | larva | 4996 | 0 |
| vir_l | larva | 6566 | 0 |
| moja_l | larva | 3970 | 0 |
| neg | negative control | 34 | 0 |

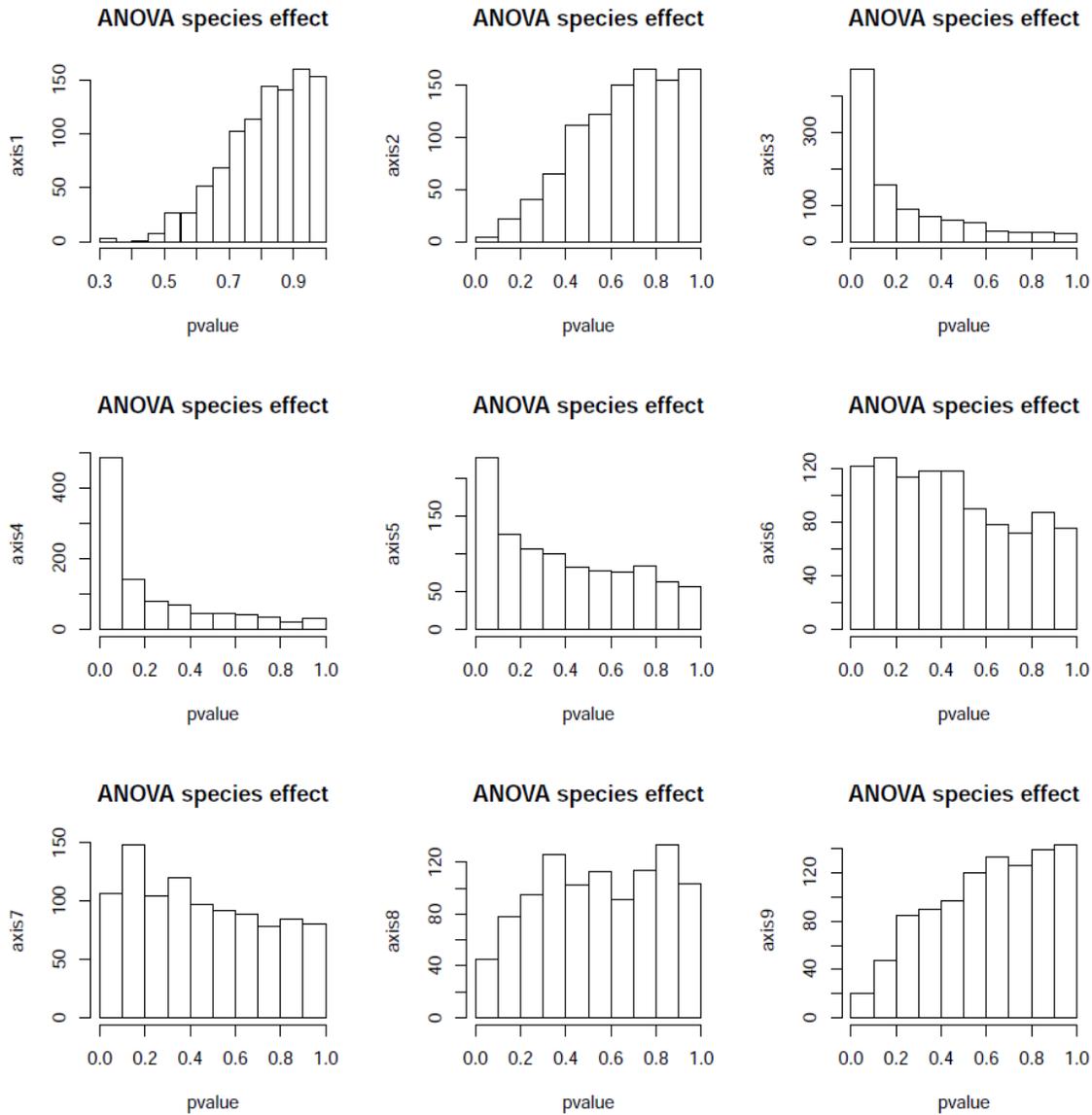

743 Supplementary Figure 1

744 P-value distributions for ANOVAs testing the alternative hypothesis that microbial communities

745 differ between wild caught *D. melanogaster* and *D. simulans* based on PCoA of Jaccard

746 distances. If there was no species effect on microbial community composition p-values are

747 expected to be uniformly distributed. PCos 1-9 are displayed. Axes 3, 4, and 5 are enriched for

748 low p-values indicating a species effect.